\documentclass[sigconf]{acmart}

\AtBeginDocument{%
  }

\usepackage{stfloats}

\copyrightyear{2025}
\acmYear{2025}
\setcopyright{cc}
\setcctype{by}
\acmConference[MUM '25]{24th International Conference on Mobile and Ubiquitous Multimedia}{December 01--04, 2025}{Enna, Italy}
\acmBooktitle{24th International Conference on Mobile and Ubiquitous Multimedia (MUM '25), December 01--04, 2025, Enna, Italy}
\acmPrice{}
\acmDOI{10.1145/3771882.3771894}
\acmISBN{979-8-4007-2015-4/25/12}
\acmISBN{978-1-4503-XXXX-X/2018/06}

\begin{document}

\title{Designing Intent Communication for Agent-Human Collaboration}


\author{Yi Li}
\orcid{0009-0009-9011-5496}
 \affiliation{%
   \institution{TU Wien}
   \city{Vienna}
   \country{Austria}}
 \email{yi.5.li@tuwien.ac.at}

\author{Francesco Chiossi}
\orcid{0000-0003-2987-7634}
\affiliation{
  \institution{LMU Munich}
  \city{Munich}
  \country{Germany}
}
\email{francesco.chiossi@ifi.lmu.de}

\author{Helena Anna Frijns}
\orcid{0000-0003-3866-7380}
\affiliation{
  \institution{Interdisciplinary Transformation University Austria (IT:U)}
  \city{Linz}
  \country{Austria}
}
\email{helena-anna.frijns@it-u.at}

\author{Jan Leusmann}
\orcid{0000-0001-9700-5868}
\affiliation{
  \institution{LMU Munich}
  \city{Munich}
  \country{Germany}
}
\email{jan.leusmann@ifi.lmu.de}

\author{Julian Rasch}
\orcid{0000-0002-9981-6952}
\affiliation{
  \institution{LMU Munich}
  \city{Munich}
  \country{Germany}
}
\email{julian.rasch@ifi.lmu.de}

\author{Robin Welsch}
\orcid{0000-0002-7255-7890}
\affiliation{%
  \institution{ Aalto University}
  \city{Helsinki}
  \postcode{02150}
  \country{Finland}}
\email{robin.welsch@aalto.fi}

\author{Philipp Wintersberger}
\orcid{0000-0001-9287-3770}
\affiliation{%
  \institution{Interdisciplinary Transformation University (IT:U)}
  \city{Linz}
  \country{Austria}}
\email{philipp.wintersberger@it-u.at}

\author{Florian Michahelles}
\orcid{0000-0003-1486-0688}
 \affiliation{%
   \institution{TU Wien}
   \city{Vienna}
   \country{Austria}}
 \email{florian.michahelles@tuwien.ac.at}

\author{Albrecht Schmidt}
\orcid{0000-0003-3890-1990}
 \affiliation{%
   \institution{LMU Munich}
   \city{Munich}
   \postcode{80337}
   \country{Germany}}
 \email{albrecht.schmidt@ifi.lmu.de}

\renewcommand{\shortauthors}{Li et al.}

\begin{abstract}
As autonomous agents, from self-driving cars to virtual assistants, become increasingly present in everyday life, safe and effective collaboration depends on human understanding of agents' intentions. Current intent communication approaches are often rigid, agent-specific, and narrowly scoped, limiting their adaptability across tasks, environments, and user preferences. A key gap remains: existing models of what to communicate are rarely linked to systematic choices of how and when to communicate, preventing the development of generalizable, multi-modal strategies. In this paper, we introduce a multidimensional design space for intent communication structured along three dimensions: Transparency (what is communicated), Abstraction (when), and Modality (how). We apply this design space to three distinct human-agent collaboration scenarios: (a) bystander interaction, (b) cooperative tasks, and (c) shared control, demonstrating its capacity to generate adaptable, scalable, and cross-domain communication strategies. By bridging the gap between intent content and communication implementation, our design space provides a foundation for designing safer, more intuitive, and more transferable agent-human interactions.

\end{abstract}




\maketitle

\section{Introduction}





Advances in artificial intelligence, sensing, and robotics have integrated intelligent agents into everyday life~\cite{straube2024robotics}. Examples range from self-driving cars~\cite{badue2021self}, cleaning robots~\cite{megalingam2025cleaning}, and delivery drones~\cite{kellermann2020drones} to virtual assistants that coordinate our schedules. As these systems become increasingly autonomous, humans must understand and predict their future actions to enable effective collaboration and seamless integration into individual and societal workflows. Without effective intent communication, even the most sophisticated agents risk being misunderstood or mistrusted, impairing collaboration, increasing cognitive load, and potentially compromising safety in critical applications. Traditional approaches to intent communication in agent-human systems face significant limitations due to their heavy reliance on context-specific designs and narrowly defined intent content. For example, while external Human-Machine Interface (eHMI) have been extensively researched in automated driving contexts \cite{dey_taming_2020}, it remains unclear whether these findings can be effectively transferred to other forms of automation, such as humanoid or household robots. This context dependency prevents these approaches from scaling across different tasks, environments, and user preferences. The core challenge lies in the absence of adaptable, generalized communication strategies that can convey intent dynamically and intuitively across diverse applications. Without such strategies, current systems remain practical only within their specific domains, creating substantial barriers to the large-scale deployment of intelligent agents across varied real-world contexts.

Previous research has explored intent communication across multiple domains, including Human-Robot Interaction (HRI)~\cite{pascher_how_2023}, autonomous vehicles~\cite{habibovic2018communicating}, and digital assistants~\cite{lair2020user}. Researchers have established unified communication models to standardize intent content between agents and humans and have classified communication modalities into visual, auditory, and haptic categories. However, a critical gap remains: the missing connection between standardized communication models, communication timing, and communication modalities. This gap prevents the formation of comprehensive guidelines for designing adaptable, multi-modal systems that achieve both robustness and accessibility.

This paper addresses this challenge by introducing a multidimensional design space for intent communication that enables designers to systematically develop communication strategies based on three key dimensions: \textbf{(1) what an agent communicates (transparency), (2) when it communicates (abstraction), and (3) how it communicates (modality)}. We demonstrate the design space's versatility and value by applying it across three distinct collaboration scenarios: bystander interaction, cooperative tasks, and shared control situations, showing how it can guide both the design and evaluation of communication strategies. Unlike existing taxonomies dependent on applications, this design space enables scalable reasoning and cross-domain applications by encouraging novel combinations of information, abstraction, and modality tailored to specific user-agent contexts. It establishes a conceptual foundation for future agent-human communication research by supporting generalization and knowledge transfer across domains. We conclude by discussing key design considerations and observed challenges to guide future improvements of this design space.

\section{Related Work}

To properly motivate our proposed study, we first define our terminology before reviewing literature on current agent-human intent communication strategies and scenarios.

In this paper, we define agents as autonomous AI systems capable of sensing, reasoning, and acting to achieve goals, mirroring core attributes of human cognition. Prior to the emergence of Large Language Models (LLMs), foundational work by Russell and Norvig~\cite{russell2016artificial} categorized AI agents into five archetypes based on their ability to learn and reason: (1) Simple reflex agents (e.g., thermostats executing condition-action rules); (2) Model-based reflex agents (e.g., robot vacuums navigating via dynamic floor plan mapping); (3) Goal-based agents (e.g., AlphaGo’s strategic gameplay); (4) Utility-based agents (e.g., autonomous vehicles optimizing safety and efficiency), and (5) Learning agents (e.g., adaptive recommendation systems like Netflix). Throughout this paper, we refer to agents generally as entities of autonomous AI systems without specifying their particular type.

Agent-Human Collaboration refers to interactive systems where autonomous agents and humans engage in coordinated activities within shared environments or toward common objectives~\cite{cila2022designing}. Drawing from Pascher et al.'s~\cite{pascher_how_2023} systematic framework, we define agent-human collaboration as encompassing three distinct interaction modalities:

\begin{itemize}
    \item Bystander: Agents and humans pursue independent goals but share a physical or digital environment (e.g., pedestrians interacting with autonomous vehicles).
    \item Cooperation: Agents and humans collaborate toward a shared goal through divided subtasks (e.g., cobots assembling products with factory workers).
    \item Shared Control: Agents and humans jointly govern a task in real time (e.g., LLM-assisted playwriting for co-creating screenplays).
\end{itemize}



\subsection{Current Intent Communication Strategies}
In human psychology, intent refers to the mental state guiding purposeful actions toward a desired goal~\cite{sheeran2002intention, chiossi2023short}. Translating this to agent-human collaboration, intent communication involves conveying an agent's purpose, anticipated actions, and internal states to users. Unlike human-to-human exchanges, where shared context and social cues facilitate understanding, agent-human intent communication hinges on system transparency and adaptability to the agent-human relationship, which shapes how users perceive and trust autonomous behaviour.

In the field of Human-Robot Interaction (HRI), Pascher et al.~\cite{pascher_how_2023} develop a comprehensive intent communication model that captures the key factors influencing intent communication strategies, derived from a scoping review of robot motion intent communication. This model identifies several critical dimensions that shape communication design: the robot type (robotic arm, humanoid, or mobile robot), the communication location (on-robot, on-world, or on-human), the intent type (motion, attention, state, or instruction), the information characteristics (spatial or temporal), and the human's role (collaborator, observer, coworker, or bystander). Once these factors are established within the communication framework, the actual communication typically occurs through various signal modalities, including auditory, visual, or haptic feedback. 

Paralleling developments in HRI, the field of eHMI research has established similar communication frameworks. Dey~\cite{dey_taming_2020} proposes a taxonomy that encompasses intent content specifications including: (1) vehicle states, such as yielding, driving, initiating movement, or remaining stationary; (2) spatial positioning of information communication, whether on the vehicle itself, integrated into infrastructure, embedded within roadways, or delivered directly to road users through their personal devices; and (3) audience configurations, targeting single receivers (unicast), multiple specific recipients (multicast), or multiple unspecified recipients (broadcast).

Despite the existence of taxonomies in HRI and eHMI for representing the design space of intent communication, previous frameworks focus primarily on communication content (i.e., what to communicate), while significant gaps remain in understanding communication modality (i.e., how to communicate) and timing (i.e., when to communicate). This limitation highlights the need to identify universal principles for designing intent communication that integrate content, modality, and timing in agent-human interaction. Xu et al.~\cite{xu_xair_2023} propose the XAIR framework for designing AI explanations in AR. The framework contains the similar problem space dimensions of when, what, and how to explain. However, their framework is AR- and explanation-specific, while ours focuses on intent communication in Human-Agent Collaboration scenarios.





In the long term, we aim to develop a unified framework for intent communication that adapts to context, ensuring intelligent agents are not just functional, but intuitive partners in human environments. This paper takes the first step towards this long-term goal by proposing a design space of intent communication to guide the exploration of such a framework.

\section{Design Space for Intent Communication}

\begin{figure*}[!t]
  \centering
  \includegraphics[width=\textwidth]{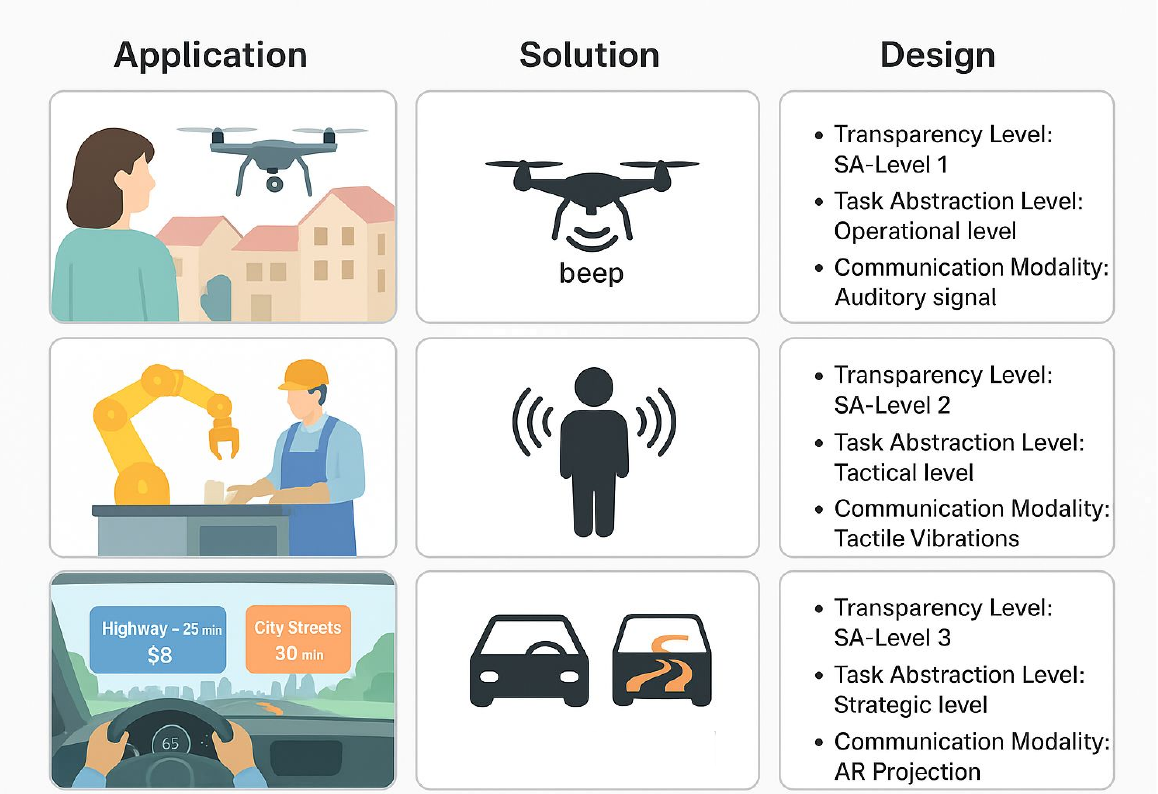}
  \caption{Illustration of three distinct application scenarios. From top to bottom: (1) Bystander Interaction, (2) Cooperative Task, and (3) Shared Control. Image generated using GPT-5.}
  \label{fig:applications}
\end{figure*}

In this section, we introduce a structured design space for intent communication in agent-human collaboration, building upon existing explicit signaling approaches~\cite{pascher_how_2023, chiossi2025designing} in the field of automation. This design space serves as a theoretical foundation for analysing, designing, and evaluating implicit communication strategies in future collaborative systems. The design space comprises three intersecting dimensions: (1) Transparency level; (2) Task abstraction level; (3) Communication Modality. Each dimension reflects a critical aspect of intent communication—what to convey, when to convey it, and how to convey it—to foster effective and trustworthy agent-human collaboration.  We display the Application, Solution and Design spaces in \autoref{fig:applications}.

\subsection{Dimension 1: Transparency Level}
The first dimension of this design space is system transparency level, which describes what internal information an agent communicates—distinct from but designed to support human Situational Awareness (SA). We adopt Endsley's three-level SA framework~\cite{endsley2017here} as a design scaffold: Transparency Level 1 exposes current system states (sensor readings, operational status) to enable human perception of the agent's condition; Transparency Level 2 reveals reasoning processes (goals, constraints, decision logic) to facilitate comprehension of why the agent acts; and Transparency Level 3 projects future intentions (planned actions, predicted outcomes) to support human anticipation. This structure treats transparency as a controllable design property~\cite{chen2014human} while acknowledging that actual SA depends on additional factors, including timing, modality, user attention, and task demands~\cite{stewart2008distributed}, among others.

For example, a warning alert notifying users of an agent's error state represents Transparency Level 1 (exposing current system condition), a verbal message from a driving assistant during bad weather conditions revealing the reasoning behind adaptive assistance represents Transparency Level 2 (explaining decision logic), while the projection of a cleaning robot's motion trajectory informing users of the agent's future plan represents Transparency Level 3 (projecting intentions)~\cite{parasuraman1997humans,lee2004trust}. While transparency alone does not guarantee SA, it provides the informational foundation necessary for SA development. Effective intent communication will therefore convey information at the appropriate transparency level to avoid both cognitive overload and undertrust of the agent\cite{hoff2015trust}.

\subsubsection{How to Analyze and Evaluate Transparency Levels}

Effective analysis and evaluation of transparency levels in human-agent systems requires multi-faceted approaches that combine real-time behavioral assessment with structured measurement techniques.

In collaborative manufacturing, SA levels are analyzed through real-time sensing of operator attention using eye and head gaze tracking~\cite{dini2017,paletta2019}. Studies demonstrate that gaze features correlate strongly with standard SA measures, such as the Situational Awareness Rating Technique (SART) and Situation Awareness Global Assessment Technique (SAGAT), and can predict handover timing and overall task performance~\cite{dini2017,paletta2019}. This enables robots to adjust their actions in synchrony with human attentional states, supporting SA-Level 1 (perception) assessment through observable attention allocation patterns.

From a measurement perspective, Schuster and Jentsch~\cite{schuster2011measurement} classified SA assessment methods into three categories: implicit behavioral indicators, subjective ratings, and direct probe techniques. They emphasize that no single measure captures SA comprehensively across all three levels, making multi-method approaches essential for accurate evaluation. Direct probe techniques (e.g., SAGAT) can assess comprehension (SA-Level 2) and projection (SA-Level 3) by querying operators' understanding of system goals and anticipated future states.

Advanced evaluation frameworks employ probabilistic belief modeling~\cite{lison2010} that integrates multimodal sensory inputs under uncertainty to maintain structured, temporally framed, and epistemically annotated representations of the robot's knowledge state. These frameworks formalize \emph{what the system knows} at each transparency level, providing the computational foundation for assessing whether the system's internal transparency design supports appropriate SA development in human operators.

Collectively, these evaluation approaches support intent communication by making explicit what the agent knows, believes, and anticipates at each transparency level. This allows for a systematic assessment of whether system transparency translates into appropriate human SA and informs adaptive communication strategies.

\subsubsection{How to Design for Different Transparency Levels}

Effective design for SA-based intent communication requires systems that adapt transparency levels to match users' current understanding and task demands. Recent research demonstrates that SA requirements in HRI are inherently dynamic, with required SA levels fluctuating based on environmental complexity, robot autonomy, and human roles~\cite{senaratne2025}. Mismatches between required and actual SA can lead to operational inefficiencies, delays, or cognitive overload, highlighting the need for user-adaptive interface design that tailors system transparency to support appropriate human SA development.

The appropriate transparency level depends significantly on users' familiarity with the task being performed. In highly familiar tasks, users possess strong mental models that enable them to infer agent intent from minimal perceptual cues alone (Transparency Level 1), recognizing what the agent will do based on observable actions. In less familiar tasks where mental models are weak or absent, higher transparency levels must be communicated explicitly, requiring systems to reveal their reasoning processes (Transparency Level 2) or project future states (Transparency Level 3) to support user comprehension and coordination. This familiarity-transparency relationship implies that adaptive systems should assess user expertise and modulate informational depth accordingly.

Task-phase considerations are critical for transparent design. Research indicates that transparency content should align with task phases~\cite{pascher_how_2023}: planning or high-uncertainty phases benefit from proactive, higher-level intent communication (Transparency Levels 2-3) that supports mental model alignment and anticipatory coordination, while execution phases require more concise, reactive cues (Transparency Level 1), particularly in time-critical contexts where excessive information may impair performance. Signaling uncertainty can improve trust calibration when presented to support decision-making while maintaining operator confidence, provided it frames uncertainty as decision support rather than system unreliability.

Design strategies should leverage established SA detection methods, including gaze-based estimation~\cite{paletta2019,dini2017}, multi-method SA measurement techniques~\cite{schuster2011measurement}, and dynamic SA frameworks~\cite{senaratne2025}, to implement adaptive transparency systems. These systems monitor human SA in real time and adjust transparency dynamically: increasing informational detail when SA drops below task requirements by providing reasoning explanations and outcome projections (Transparency Levels 2-3), while reducing detail when SA is sufficient to prevent cognitive overload. This closed-loop approach ensures that system transparency supports appropriate SA development without overwhelming users.

To summarize, we recommend implementing adaptive transparency levels that: 

\begin{itemize}
    \item Modulate intent information granularity in real-time based on SA indicators and task phase,
    \item Prioritize essential information during high-tempo situations to prevent cognitive overload,
    \item Include uncertainty cues when relevant to decision-making and risk assessment, and
    \item Calibrate transparency depth as operator expertise and mental models mature.
\end{itemize}

This design approach ensures intent communication remains contextually relevant and cognitively efficient, directly supporting SA maintenance and effective agent-human collaboration.

\subsection{Dimension 2: Task Abstraction Level} 
The second dimension of this design space is the task abstraction level, which concerns the temporal and conceptual abstraction of the task at hand. This dimension includes operational, tactical, and strategic levels, as commonly found in both cognitive models and robotics planning literature~\cite{chen2014human,wintersberger2020automated}. The operational level corresponds to immediate, concrete actions; the tactical level represents short-term planning and sequencing; and the strategic level reflects broader goals and adaptive reasoning. These three levels capture different temporal aspects of intent. By combining them with the three SA levels, we can categorize context-specific intent content based on what information is needed and when. 

\subsubsection{How to Analyze and Evaluate Task Abstraction Levels.}
Systematic analysis of task abstraction levels requires examining the temporal and conceptual complexity inherent in different HRI domains. Research demonstrates that HRI tasks can be categorized into three primary domains with distinct abstraction requirements: navigation~\cite{watanabe2015communicating, koay2013hey}, object manipulation~\cite{newbury2022visualizingrobotintentobject, anderson2016projecting}, and collaborative interactions~\cite{hoffman2004collaboration}. Navigation tasks involve robots communicating movement intentions within shared environments, ranging from street crossings~\cite{watanabe2015communicating} to guided assistance~\cite{koay2013hey} and industrial space navigation~\cite{chadalavada2015thats}. These tasks primarily operate at operational and tactical levels, focusing on immediate spatial decisions and short-term path planning. Object manipulation tasks center on communicating handling intentions, for example, grasping~\cite{newbury2022visualizingrobotintentobject}, handover coordination~\cite{newbury2022visualizingrobotintentobject}, and item retrieval~\cite{nicolescu2005task, leusmann2025understanding}, which typically requires operational-level precision with tactical-level sequencing. Collaborative interaction tasks integrate multiple communication modes to achieve complex shared goals, such as assembly operations in cooperative or shared control tasks, demanding communication across all abstraction levels to convey immediate actions, subtask coordination, and strategic objectives.

\subsubsection{How to Design for Different Task Abstraction Levels.}
Effective design requires systematic alignment of intent communication with the temporal and conceptual abstraction demands of specific tasks. This alignment ensures information is conveyed with appropriate granularity, supporting human understanding and coordination while avoiding cognitive overload. 

Researchers should begin by systematically analyzing their target task to identify required abstraction levels and decompose communication requirements into distinct components: 
\begin{itemize}
    \item Operational communication addresses immediate, concrete actions the agent will perform (e.g., initiating movement, grasping objects);
    \item Tactical communication conveys short-term plans or action sequences linking operational steps toward near-term goals (e.g., planned path segments, ordered assembly steps); and
    \item Strategic communication encompasses long-term goals, overarching objectives, and adaptive reasoning guiding overall task direction.
\end{itemize}

After defining the sub-components, researchers should ensure internal consistency within each abstraction level by avoiding the blending of details from different levels within a single communication cue. This approach maintains clear conceptual boundaries and enables smooth transitions between abstraction levels as tasks progress.

\subsection{Dimension 3: Communication Modality}
The third dimension of the proposed design space is communication modality, which addresses the sensory channel through which intent is conveyed: visual, auditory, haptic, and potentially more speculative modalities such as smell and temperature. Common modalities differ significantly in accessibility, expressiveness, and environmental suitability. Visual modalities (lights, displays, gestures) offer spatial precision but can be ineffective in cluttered spaces~\cite{ganesan2018better}, while auditory signals (speech, alarms) can cut through visual noise but may be drowned out in loud settings or cause irritation~\cite{orthmann2023sounding}. Haptic feedback (vibrations, force feedback) provides discrete, localized prompts but is often less expressive~\cite{chiossi2022design}. Multimodal communication, combining channels either simultaneously or sequentially, can improve robustness and reduce cognitive effort through redundancy, yet may also lead to sensory overload or conflicting signals~\cite{sarter2006multimodal, fox2021relationship}. The most effective intent communication results from matching modality choices to the context, the user's state, and the task demands.

\subsubsection{How to Analyze and Evaluate Communication Modality}
Systematic analysis of communication modality requires both systematic classification frameworks and domain-specific investigations. Comprehensive modality analysis employs established classification frameworks that capture the breadth of communicative channels. Feine et al.~\cite{feine2019taxonomy} provide a validated taxonomy for conversational agents, categorizing verbal, visual, auditory, and invisible cues into ten subcategories grounded in interpersonal communication theory and validated through empirical methods, offering a systematic framework for evaluating social communicative elements. Applied studies demonstrate domain-specific approaches to modality evaluation. Industrial contexts require mapping human-cobot scenarios to specific communicative modes including proxemics, kinesics, haptics, speech, and visual signals~\cite{gross2024communicative}, while automated driving research employs comprehensive taxonomies that systematically code visual, auditory, and haptic concepts to characterize vehicle intent signaling effectiveness~\cite{dey2020taming}. These domain-specific frameworks enable researchers to systematically evaluate which modality combinations are most effective for particular interaction contexts and task requirements. 

This research demonstrates that communication modalities represent strategic design choices rather than arbitrary selections. Whether formalized through taxonomies or explored in applied settings, studies reveal how different channels, including unintended cues, influence intent reception. Across domains, modality functions as a central factor in human interpretation of agent actions, making systematic evaluation essential for effective intent communication design.

\subsubsection{How to Configure Communication Modality}
Effective modality configuration requires Situated Modality Coordination: the deliberate combination and calibration of communication channels to match environmental conditions, task demands, and user cognitive states while avoiding sensory overload or conflicting signals. This approach recognizes that both deliberate signals and incidental cues~\cite{hegel2011typology} influence user interpretation, requiring coherent orchestration across modalities to maintain intent clarity.

Drawing from established frameworks~\cite{feine_taxonomy_2019, holthaus2023communicative, dey2020taming}, modality configuration should consider signal origin, deliberateness, and clarity alongside environmental and task constraints. Domain-specific research~\cite{gross2024communicative, shiokawa2025beyond} demonstrates that modality selection must be driven by communicative purpose within specific collaboration contexts rather than fixed mappings. Effective Situated Modality Coordination involves: 

\begin{itemize}
    \item Mapping modality choices to environmental constraints (noise levels, visual clutter) to ensure signal detectability;
    \item Leveraging complementary channel strengths, for example, haptics for immediate localized alerts, visuals for persistent spatial information;
    \item Integrating deliberate and incidental cues so unintended signals (motor sounds, idle motions) reinforce rather than contradict intended messages;
    \item Sequencing modalities to manage attention by introducing high-salience cues first followed by detailed information through slower channels;
    \item Empirically testing multimodal configurations to detect conflicts or cognitive overload before deployment; and
    \item Adjusting modality emphasis based on task phases or user states while maintaining predictable patterns for user orientation.
\end{itemize}

Properly configured modalities should form a unified communicative act that matches interaction context, fosters shared awareness, minimizes ambiguity, and facilitates effective agent-human collaboration through strategically coordinated sensory channels.

\section{Design Space Application}
To demonstrate the practical value of the proposed three-axis design space for intent communication in human–agent collaboration, we applied it to three distinct human-agent collaboration scenarios: bystander interaction, cooperative tasks, and shared control. The design space is conceived as a cube whose axes represent, first, the depth of information disclosed according to the transparency hierarchy; second, the abstraction level of the task in question—operational, tactical, or strategic; and third, the communication modality, namely visual, auditory, or haptic. Treating each intent message as a point inside this design space enables systematic reasoning about what the agent reveals, when it does so, and through which sensory channel.

The first application involves a bystander scenario where a delivery drone operates in a residential neighborhood for daytime delivery mission. As the drone descends toward a designated yard location to drop off a package, it emits a brief but distinctive auditory beep at regular intervals to alert nearby pedestrians and residents of its immediate presence and ongoing delivery activity. Here the cue reveals the drone's basic perception of the drop-off location (SA-Level 1) at an operational level via auditory signals. This auditory communication is particularly effective in public residential spaces where visual attention may be divided, allowing pedestrians to quickly recognize the drone's presence, thereby supporting safe coexistence in shared airspace.

The second application concerns a cooperation scenario where a collaborative robot assists a human in an industrial drilling task. As the robot approaches completion of the alignment phase and prepares for synchronous motion with the human operator, it delivers tactile vibrations through wearable devices worn by the worker, signaling readiness for coordinated action. Here the cue exposes the robot's comprehension of the joint task state and timing requirements (SA-Level 2) at a tactical level with haptic feedback to convey operational readiness. This haptic communication allows the human operator to understand the current progress of the alignment task and anticipate the transition to the next phase of coordinated drilling, providing a communication method that is both private to the individual operator and robust against the ambient noise typical of industrial workshops.

The third application features a shared control scenario where an autonomous vehicle suggests a highway route to the airport under time pressure, but the driver prefers taking city streets. Before the trip begins, the vehicle projects its proposed route onto the roadway using augmented reality display technology, showing side-by-side route options with clear time and cost trade-offs for each alternative. Here the cue projects the vehicle's future planning decisions and strategic reasoning for route selection (SA-Level 3) at a strategic level through visual projection signals. This visual communication enables the driver to evaluate route options, allowing them to accept the vehicle's recommendation or override the autonomous system's suggestion, maintaining human authority while benefiting from the vehicle's advanced sensing and planning capabilities.

As these examples occupy distinct cells of the design space, they illustrate how designers can consciously distribute communicative effort across space, time, and modality rather than defaulting to a single channel or information tier.  Moreover, the exercise exposes sparsely populated regions—strategic-level haptics, for instance—that may hold untapped potential in supervisory contexts or environments with constrained sensory channels.

\begin{table*}[!t]
  \centering
  \caption{Critical considerations when implementing the design space}
  \resizebox{\textwidth}{!}{%
    \begin{tabular}{p{0.26\textwidth} p{0.72\textwidth}}
      \toprule
      \textbf{Consideration} & \textbf{Impact for Designers} \\ \midrule
      Trust vs.\ Cognitive Load & Increase transparency only to the point it enhances calibrated trust; beyond that threshold, extra detail becomes distracting and counter-productive. \\
      Temporal Coordination & Align the timing of SA-Level cues with users’ decision windows: deliver SA-Level 1 in real time, defer SA-Level 3 until cognitive lulls. \\
      Ergonomics and Accessibility & Choose modalities that suit the sensory context and combine channels to offset visual, auditory, or haptic limitations. \\
      Generalizability & Tag intent messages solely by SA-Level depth, task horizon, and modality to create reusable chunks that transfer across domains. \\
      Explicit vs.\ Implicit Signalling & Balance overt notifications (e.g., speech) with subtler motion or light patterns to maintain task flow while preserving interpretability. \\
      Cultural \& Ethical Adaptability & Tailor colour semantics, privacy levels, and autonomy cues to local norms; use the design space as a scaffold for responsible adaptation. \\
      \bottomrule
    \end{tabular}%
  }
  \label{tab:considerations}
\end{table*}


\subsection{Analytical Benefits and Design Implications}
\label{subsec:benefits}

Mapping a product’s intent messages often reveals overreliance on particular modalities, such as vision, or neglect of deeper SA levels. For example, automated vehicles frequently use visual eHMI like displays or lights as the primary intent communication channel, despite evidence that such reliance can limit accessibility and robustness, especially in visually cluttered or low-visibility environments~\cite{tekkesinoglu2025advancing, roche2019behavioral}. Moreover, research in multimodal interaction consistently finds that systems relying on a single modality, especially vision, risk exclusion of users in diverse real-world contexts and miss opportunities for more robust, redundant, and user-centered communication strategies~\cite{hu2025vision, dritsas2025multimodal}.

The geometric view, therefore, supports more principled trade-offs between informativeness and cognitive economy.  
Because intent “chunks’’ are labeled solely by their coordinates in the design space, designers can also transplant successful combinations across domains: an SA-Level 1–auditory cue that excels in a warehouse frequently transfers to healthcare with minimal adaptation, while SA-Level 3–visual projections suit strategic planning tasks in comparatively static settings. Empty regions of the design space flag research opportunities where novel combinations could address unmet user needs.

Moreover, the model encourages generalization: by abstracting communication strategies from specific use cases, it becomes possible to derive cross-domain principles. For instance, SA-Level 1–auditory combinations may consistently excel in dynamic environments, while SA-Level 3–visual cues may be more suited to strategic planning in static settings.

\section{Design Considerations and Observed Challenges}

Applying the design space across the three scenarios and several pilot deployments unearthed six recurring considerations that merit attention during implementation, which we summarize in \autoref{tab:considerations}.  

First, the tension between transparency and cognitive load remains acute. Rich explanations undoubtedly foster calibrated trust, yet higher-level SA-Level disclosures can overwhelm attention if delivered too often or at inopportune moments. Judicious pacing is therefore essential.  

Second, task abstraction is as critical as transparency. SA-Level 1 cues typically require delivery within milliseconds of an impending action, whereas SA-Level 3 messages can be deferred until natural cognitive lulls, thereby avoiding premature or stale updates. 

Third, the modality axis foregrounds ergonomic and accessibility concerns. Visual messages excel when sightliness are clear, auditory signals dominate public but visually cluttered spaces, and haptics prove invaluable when both sight and hearing are compromised. Multimodal blends can further mitigate the limitations of any single channel.  

Fourth, the design space’s abstract structure promotes generalisability. When tagging intent messages only with SA-Level  depth, task horizon, and modality, engineers can implement open libraries that can be used from factory floors to autonomous vehicles without extensive rewrites.  

Fifth, the design space accommodates both explicit and implicit signaling. Designers may layer overt speech with subtler kinematic or light cues, balancing interpretability against continuity of task flow.  

Finally, cultural and ethical variation cannot be ignored. Colour semantics, privacy norms, and expectations of autonomy differ across regions and sectors; the design space does not dictate solutions but offers a structured basis for reasoning about adaptation and compliance.

Overall, the design space enables systematic reasoning about intent communication strategies in terms of informational depth, temporal dynamics, modality selection, and contextual fit. By making explicit the dimensions along which communication design can be evaluated and optimized, the design space advances toward a more principled, adaptable, and user-centered approach to agent-human collaboration.

\section{Conclusion}
This paper introduces a multidimensional design space that systematically connects the three fundamental dimensions of intent communication: what agents communicate (transparency), when they communicate (abstraction), and how they communicate (modality). For each dimension, we provide guidelines to analyze, evaluate, and design the current factor to better practice intent communication. We conceptualize intent as modular components that occupy distinct positions within this design space, providing a practical foundation for developing and evaluating transparent agent behaviors across domains. Applied across three distinct collaboration scenarios, this design space enables a more systematic approach to designing intent communication strategies across agent-human systems. We consider this design space as a generative foundation for ongoing research, evaluation, and tool development. Our long-term goal is to develop a unified framework that supports transparent, trustworthy, and adaptive collaboration between humans and intelligent agents across the expanding landscape of agent-human interaction.

\begin{acks}
This research was supported by the Deutsche Forschungsgemeinschaft (DFG, German Research Foundation) under project number 529719707, "Multimodale Kommunikation von Intention in autonomen Systemen", and by the Austrian Science Fund (FWF) under project number I6682-N as part of the joint DFG–FWF collaboration on multimodal intention communication in autonomous systems.
\end{acks}

\bibliographystyle{ACM-Reference-Format}
\bibliography{Reference}

\appendix

\end{document}